\documentstyle[prl,twocolumn,aps]{revtex}

\tighten
\begin{document}
\draft
\twocolumn[\hsize\textwidth\columnwidth\hsize\csname @twocolumnfalse\endcsname

\title{Quantum searching with continuous variables}
\author{Arun K. Pati$^{1,2}$, Samuel L. Braunstein$^{1}$ and Seth Lloyd$^{3}$}
\address{$^{1}$Informatics, University of Wales, Bangor LL57 1UT, UK}
\address{$^{2}$Theoretical Physics Division, BARC, Mumbai - 400 085, INDIA.}
\address{$^{3}$MIT Department of Mechanical Engineering,
MIT 3-160, Cambridge, Massachusetts 02139, USA}

\date{\today}
\maketitle

\begin{abstract}
A fast quantum search algorithm for continuous variables is presented. The 
result is the quantum continuous variable analog of Grover's algorithm 
originally proposed for qubits. A continuous variable analog of the 
Hadamard ({\it i.e.}, Fourier transform) operation is used in conjunction 
with inversion about the average of quantum states to allow the approximate 
identification of an unknown quantum state in a way that gives a 
square-root speed-up over search algorithms using classical continuous 
variables. Also, we show that this quantum search algorithm is robust for 
a generalised Fourier transformation on continuous variables.
\end{abstract}
\vspace{3ex}
]

Quantum systems can register and process information in ways that classical 
systems cannot. As a result, it is possible for quantum computers to 
perform certain computational tasks faster than any classical computer 
\cite{pb,rf,dd,dj,bv,drs,sl,div,sl1}. It is becoming increasingly clear
that at the heart of quantum computation lies two basic quantum phenomena,
one is quantum interference and the other quantum entanglement. The real 
upsurge of interest in this field came after Shor's \cite{ps} remarkable
discovery of an algorithm for factoring large numbers \cite{ej}. 
Subsequently, a fast quantum search algorithm has been discovered by 
Grover \cite{lg}, which takes $O(\sqrt N)$ steps instead of $N$ steps to 
search an unmarked item in a unsorted list of $N$ entries. Later the 
optimality of this search algorithm was proved \cite{bt,za}. In particular,
it was shown that this search algorithm can use almost any unitary 
transformation on qubit states \cite{lgr}. Further, it has even been argued 
that this algorithm would work even without entanglement (though at a cost 
in resources) \cite{sll}. 

These algorithms are usually implemented on quantum systems whose observables 
have discrete spectra, such as a collection of two-level atoms, ions, or 
spin-$\case{1}/{2}$ particles (called qubits). However, there are other 
classes of quantum systems whose observables form continuous spectra. So, 
it is important to know how these algorithms can be generalised for 
continuous quantum variables? With the recent advances in our ability to
manipulate continuous quantum information in teleportation \cite{sb},
error-correction codes \cite{slb,sl2} and its feasibility of implementation 
using linear devices \cite{sam}, it is natural to ask whether one can 
provide some quantum algorithms that might be implemented on a continuous 
variable quantum computer. In fact, the usefulness of quantum computation 
over continuous variables has been recently emphasised \cite{lb}. It has 
been shown that universal quantum computation over continuous variable
is not only possible, but could be effected using simple non-linear 
operations with coupling provided solely by linear operations \cite{lb}.
These operations form  a universal set of quantum gates for continuous 
variables allowing `quantum floating point' arithmetic. While discrete 
quantum computation can be thought of as the coherent manipulation of 
qubits, continuous quantum computation can be thought of as the manipulation 
of `qunats', where the qunat (pronounced as `Q nat') is the unit of 
continuous quantum information.

In this letter, we propose a fast quantum search algorithm with continuous
variables. Here a continuous variable can be anything, {\it e.g.}, position,
momentum, energy (unbounded) or amplitudes of the electromagnetic field.
With the help of the Fourier transform (viewed as an active operation) on
a continuous basis state (analogous to the Hadamard transform in the case 
of qubits) and a suitably generalised inversion operator, we construct a 
search operator which can be implemented on a continuous variable quantum 
computer. The inversion operator requires the projection operator for 
continuous basis states which we discuss. We show that the application of 
the compound operator takes $O(\sqrt N)$ iterations to search an unmarked
item in a list of $N$ entries. Further, we generalise our quantum searching
with continuous variables to one based on a generalised Fourier 
transformation and which still gives a square root reduction in the
number of steps. This shows that the quantum search algorithm with continuous
variables is robust to the choice of arbitrary Fourier transformations. 
We also discuss the robustness of the search algorithm if one uses strongly 
peaked normalisable states instead of ideal infinite-energy position 
eigenstates.

Here, we discuss how to perform a quantum  search  algorithm using 
continuous variables. First we need to map a conventional discrete search 
problem into a continuous variable context. Suppose we have a function 
$f(k): K \rightarrow \{0,1\}$ defined on a domain $K$ with
$k \in  K=\{1,2,\ldots,N\}$. This function has a non-zero value equal to 
$1$ for some element $k = k_f$ and is $0$  for all other elements in the set 
$K$. Our task is to discover the value of $ k_f$ given the ability to apply 
the function $f$ to inputs or superpositions of inputs, and given no 
further information about the function $f(k)$. In order to implement
this in a quantum computer with {\it continuous variables} we require
a collection  of $n$ qunats. The state vector of each qunat belongs to a 
Hilbert space of infinite dimension. Since we have an infinite number of 
basis states, we cannot map each basis state within the Hilbert space 
onto each entry in the set $K$. (This would be a many-to-one mapping.)
One could avoid this problem by choosing a subspace of the full Hilbert 
space with $N$ disjoint regions of the spectrum. However, we do not discuss
this approach in detail here.

Let us consider a collection of $n$ continuous variables whose Hilbert 
space is spanned by a basis of states 
$\arrowvert x \rangle = \arrowvert x_1,x_2, \ldots ,x_n \rangle$,
satisfying the orthogonality condition 
$\langle x \arrowvert x' \rangle = \delta(x_1 - x_1')\cdots
\delta(x_n - x_n') = \delta(x - x')$.
For example, one can consider a compact region of the state space divided
into $N$ equal subvolumes, each with measure $\Delta x^n$, one for each member of the set 
$K$. Let $x_f$ be the centre of the subvolume corresponding to $k_f$. In 
the context of this continuous variable embedding, executing the function 
$f$ corresponds to adjoining an extra state to the system, originally in the 
state $\arrowvert 0 \rangle$, and applying an operator 
$U_f: \arrowvert x \rangle \arrowvert 0 \rangle \rightarrow \arrowvert 
x \rangle \arrowvert 1 \rangle$ if $x$ belongs to the region corresponding 
to $k_f$ and
$\arrowvert x \rangle \arrowvert 0 \rangle  \rightarrow
\arrowvert x \rangle \arrowvert 0 \rangle$ 
otherwise. Clearly, if one samples the region at random by applying the 
operator to a series of random points, it will take $O(N)$ calls of the 
operator to find $k_f$.

If one exploits the power of quantum superposition and entanglement, 
however, fewer function calls are required. 
Our approach is discussed below. Let us pick an initial state 
in the position basis such as
$\arrowvert x_i \rangle = \arrowvert x_1,x_2,\ldots,x_n \rangle_i$
for a quantum computer with continuous spectrum at random. The final 
(target) state is given by 
$\arrowvert x_f \rangle = \arrowvert x_1,x_2,\ldots,x_n \rangle_f$.
We need a suitable unitary operator, which can take the initial state to 
the final state. Just as we have the Hadamard transformation in discrete 
computation, one of the basic operations with continuous variables is the
Fourier transformation between position and momentum variables in phase 
space. By defining the Fourier transformation as an active operation on 
$n$ qunat states $\arrowvert x \rangle$ we can write it as
\begin{equation}
{\cal F} \arrowvert x \rangle = {1 \over \sqrt{\pi^n} } \int dy\, e^{2ixy} 
\arrowvert y \rangle\;,
\end{equation}
where $xy = x_1y_1 + \cdots+x_ny_n$, $\arrowvert y \rangle =
\arrowvert y_1,y_2,\ldots,y_n \rangle$ and both $x$ and $y$ are in the 
position basis. This has been used by one of the present authors 
\cite{slb,sam} in developing an error correction code for continuous 
variables. This Fourier transformation can be straightforwardly applied 
in physical situations. For example, when $\arrowvert x \rangle$ represents 
quadrature eigenstate of a set of modes of the electromagnetic field, 
${\cal F}\arrowvert x \rangle$ is simply an eigenstate of the conjugate 
quadrature.

Suppose, we apply the unitary operator $\cal F$ to a basis state 
$\arrowvert x_i \rangle$, then the relative amplitude of finding the system 
in the target state $\arrowvert x_f \rangle$ is
$\langle x_f \arrowvert {\cal F} \arrowvert x_i \rangle = {\cal F}_{fi} =
e^{2ix_i x_f}/\sqrt{\pi^{n}}$. Therefore, the relative probability of finding 
the system in the final qunat states will  be given by 
$\arrowvert {\cal F}_{fi} \arrowvert^2= 1 /\pi^{n} $. Hence, we have to
repeat the experiment at least 
$1/ \arrowvert {\cal F}_{fi} \arrowvert^2 = \pi^n $
times to successfully obtain the state $|x_f \rangle$. Here, we prove 
that search algorithm based on continuous variable can take $\sqrt {\pi^n} $ 
steps to reach the final state starting from an initial state. (Here, we 
may identify the number of entries $N$ with $\pi^n$.)

The next operator we need is the unitary operator, which can invert the 
sign of a basis state $\arrowvert x \rangle$. We can define the selective 
inversion operator for a continuous basis $\arrowvert x \rangle$ as 
\begin{equation}
I_x =  1- 2 P_{\Delta x} \;,
\end{equation}
where  $ P_{\Delta x} $ is the projection operator for continuous
variables. Unlike the discrete case we cannot define the projection 
operator for the basis 
$\arrowvert x \rangle$ as $P_x=\arrowvert x \rangle \langle x \arrowvert$, 
because the operator $P_x$ is an ill defined and it will not satisfy 
$P_x^2 = P_x$. The correct projection operator for continuous variables 
is defined \cite{cohen} as 
\begin{equation}
P_{\Delta x}  = \int_{x_0 - {\Delta x /  2}}^{x_0 + {\Delta x / 2}} 
dx'\arrowvert x' \rangle \langle x' \arrowvert\;.
\end{equation}
The reason for this definition is that we cannot project an arbitrary
state which is represented in terms of continuous basis state onto a point 
to get the exact eigenvalue. There will be always a spread within an 
interval. We can only project a state around $x_0$ to a selectivity 
$\Delta x$ of the measuring apparatus. It is not possible to design a 
device to make a perfectly selective measurement of a continuous variable. 
The interval $[x_1, x_2]$ cannot be narrowed down, because it will always 
contains an infinite number of eigenvalues \cite{cohen}. Thus, if we have 
a wave packet the effect of projection is to truncate it around $x_0$ within 
an interval $\Delta x $. This operator satisfies 
$P_{\Delta x}^2 = P_{\Delta x}$ and 
$P_{\Delta x} \arrowvert x \rangle = \arrowvert x \rangle$ as expected. 
With the help of the above inversion operator we can construct a compound 
search operator $\cal C$ defined as
\begin{equation}
{\cal C} = - I_{x_i}\, {\cal F}^{\dagger} I_{x_f} {\cal F}\;.
\end{equation}
It may be remarked that the selective inversion of the target state
$\arrowvert x_f \rangle$ can be achieved by attaching an ancilla qunat 
and considering the quantum XOR circuit for continuous variables \cite{slb}. 
If a quantum circuit exists that transforms 
$\arrowvert x \rangle \arrowvert a \rangle \rightarrow
\arrowvert x \rangle \arrowvert f(x)+ a \rangle$, then by choosing the 
ancilla state
$\arrowvert a \rangle = {\cal F} \arrowvert {\pi}/{2} \rangle = 
\int dy\, e^{i \pi y} \arrowvert y \rangle/\sqrt{\pi^n}$
we can selectively invert the state
$\arrowvert x \rangle$ for which $f(x) =1$, {\it i.e.},
$\arrowvert x \rangle  {\cal F} \arrowvert {\pi}/{2} \rangle \rightarrow
-\arrowvert x \rangle {\cal F} \arrowvert {\pi}/{2} \rangle$.

Let us define a state $\arrowvert \tilde{x}_f \rangle\equiv 
{\cal F}^{\dagger} \arrowvert x_f \rangle$.
We can show that the operator ${\cal C}$ can preserve the two-dimensional
subspace spanned by the states $\arrowvert x_i \rangle$ and 
$\arrowvert \tilde{x}_f \rangle$. First, we show the action
of ${\cal C}$ on $\arrowvert x_i \rangle$. This can be expressed as
\begin{eqnarray}
{\cal C} \arrowvert x_i \rangle = \arrowvert x_i \rangle - 
4 P_{\Delta x_i} {\cal F}^{\dagger}
P_{\Delta x_f} {\cal F} \arrowvert x_i \rangle + 2 {\cal F}^{\dagger}
P_{\Delta x_f} {\cal F} \arrowvert x_i \rangle\;, \!\!
\end{eqnarray}
where $P_{\Delta x_i} =
 \int_{x_{i1}}^{x_{i2}} dx_i'\,\arrowvert x' \rangle_i 
{_i}\langle x' \arrowvert$, ($x_{i1} = x_0 - {\Delta x_i /2}$, 
$x_{i2} = x_0 +  {\Delta x_i / 2}$ ) and likewise for $P_{\Delta x_f}$. 
Using these facts we can simplify the above equation to
\begin{eqnarray}
{\cal C} \arrowvert x_i \rangle = (1 - {4 \over \pi^n}) 
\arrowvert x_i \rangle + {2 \over \sqrt{\pi^n}} \int_{x_{f1}}^{x_{f2}} \!
dx_f' \,e^{2ix_i x_f'} {\cal F}^{\dagger}
 \arrowvert x_f '\rangle\;.\!\!
\end{eqnarray}
Similarly, we can evaluate the action of ${\cal C}$ on 
$\arrowvert \tilde{x}_f \rangle$. It is given by 
\begin{eqnarray}
{\cal C} \arrowvert \tilde{x}_f \rangle = \arrowvert \tilde{x}_f \rangle
- {2 \over \sqrt{\pi^n}} \int_{x_{i1}}^{x_{i2}} dx_i'\,e^{2ix_i' x_f }
 \arrowvert x_i '\rangle\;.
\end{eqnarray}
Thus, the operator ${\cal C}$ creates superpositions of two qunat states
just as Grover's operator creates superpositions of two qubit states. Once
we understand the action of ${\cal C}$ on qunats
we can obtain the total number of steps required in 
reaching the target state. Here, we use geometric structures from the
projective Hilbert space of a quantum system to obtain the number of steps
in the quantum searching.
The projective Hilbert space admits a natural measure of distance called 
Fubini-Study distance \cite{ap}. This measures the shortest distance between 
any two (not necessarily normalized) states $\arrowvert \psi_1 \rangle$ and 
$\arrowvert \psi_2\rangle$ whose projections on $\cal P$ are 
$\Pi(\psi_1)$ and $\Pi(\psi_2)$, respectively. This can be defined as 
\begin{equation}
d^2( \arrowvert \psi_1 \rangle,\arrowvert \psi_2 \rangle) = 4 \biggl(
1 - \bigg{\arrowvert} \bigg{\langle}
{ \psi_1 \over  \arrowvert\arrowvert\psi_1 \arrowvert\arrowvert}
\bigg{\arrowvert}
{ \psi_2  \over \arrowvert \arrowvert \psi_2 \arrowvert \arrowvert}
\bigg{\rangle}  \bigg{\arrowvert}^2 \biggr).
\end{equation}
Here, the vectors $\arrowvert\psi_1\rangle$ and $\arrowvert\psi_2\rangle$
can be quantum states over continuous variables or over discrete variables. 
For unnormalisable states the above definition still works provided it is 
understood that the norm of the states can be made finite. In dealing with 
position eigenstates we can imagine that either the particle is moving in 
a finite space \cite{cohen} so that position eigenstates do not diverge or 
one can use normalisable states having strong peaks around some value of 
the position axis.

During the quantum searching with continuous variables we  want
to  reach  a state $\arrowvert \tilde{x}_f \rangle$ from an initial state
$\arrowvert x_i \rangle$. This means we  have  to travel a shortest distance
between  these  states which is given  by  
$d^2(\arrowvert x_i \rangle,\arrowvert \tilde{x}_f \rangle) 
= 4(1 -  {1 / \pi^n})$.
One application of the operator ${\cal C}$ creates a state
$\arrowvert x_i \rangle^{(1)} = {\cal C} \arrowvert x_i \rangle $.
We calculate the shortest distance  between  the  resulting  state
$\arrowvert x_i \rangle^{(1)} $ and the initial state 
$\arrowvert x_i \rangle $. We note that the overlap of these states is 
given by
$\langle x_i \arrowvert {\cal C} \arrowvert x_i \rangle 
=(1-{4/\pi^n} ) \langle x_i \arrowvert x_i \rangle +2\Delta x_f/\pi^n$.
For large database search $N=\pi^n$ is very large and if we assume that the
measuring device has a narrow selectivity, then $\Delta x_f$ is also small.
Hence, we can neglect the second term in the overlap (as it is a product of 
two small terms). Also note that term $\langle x_i \arrowvert x_i \rangle$ 
is not normalised but nevertheless it cancels out in during calculation.
With this idea in mind we can evaluate the shortest distance between these
states which is given by
$d^2(\arrowvert x_i \rangle,\arrowvert x_i^{(1)} \rangle) \approx32/\pi^n$. 
Thus in one application of the search operator ${\cal C}$ we can move the 
initial basis a shortest distance $O({1 / \sqrt{ \pi^n } })$. Therefore, 
to travel the full distance on the quantum state space we need $N_s$ number 
of steps given by
\begin{equation}
N_s  = {d(\arrowvert x_i \rangle,\arrowvert \tilde{x}_f \rangle )  
\over d(\arrowvert x_i \rangle,\arrowvert x_i^{(1)} \rangle) } \approx
O(\sqrt {\pi^n})\;.
\end{equation}
This shows that a quantum computer based on qunats can take
$O(\sqrt {\pi^n} )$ applications of ${\cal C}$ to reach the target state
which otherwise would have taken $O(\pi^n)$ number of steps by the
application of ${\cal F}$ on $\arrowvert x_i \rangle$. Because the state is
moving along a geodesic each application of ${\cal C}$ rotates the initial
state in the right direction. This is our quantum search algorithm
with continuous variables.

Instead of position eigenkets one can use strongly peaked normalisable 
state such as 
\begin{equation}
\arrowvert r_i \rangle = {1\over (2 \pi \epsilon)^{n/4}}
\int dx\, \exp[- {{(x -x_i)^2 }\over{4 \epsilon^2}} ] 
\arrowvert x \rangle \;.
\end{equation}
When $\epsilon \rightarrow 0$, the state $\arrowvert r_i \rangle$ becomes 
a position eigenstate $\arrowvert x_i \rangle$. Our algorithm
can be practically implemented with such states. One can see that the
action of the search operator ${\cal C}$ on $\arrowvert r_i \rangle$ gives
\begin{eqnarray}
{\cal C} \arrowvert r_i \rangle &=& 
(1 - {4 \over \pi^n}) \arrowvert r_i \rangle + 
{2 \over \sqrt{\pi^n}} \frac{1}{(2 \pi \epsilon)^{n/4}} \nonumber \\
&&\times\! \int dx\int_{x_{f1}}^{x_{f2}} dx_f'
 e^{-{(x -x_i)^2 }/4\epsilon^2+2ix x_f'} 
{\cal F}^{\dagger}  \arrowvert x_f '\rangle\;.\!\!
\end{eqnarray}
From the above formula one can see that a single application of search
operator moves the initial state $\arrowvert r_i \rangle$ a distance given by
$d^2(\arrowvert r_i \rangle,\arrowvert r_i^{(1)} \rangle) \approx
{48 / \pi^n}$. Now, if one defines the target state as
\begin{equation}
\arrowvert r_f \rangle =  
\frac{1}{(2 \pi \epsilon)^{n/4}} \int dx 
\exp[- \frac{(x -x_f)^2 }{4 \epsilon^2} ]
\arrowvert x \rangle\;,
\end{equation}
then one can check that the total (shortest) distance between the initial 
state $\arrowvert r_i \rangle$ and the desired state 
$\arrowvert {\tilde r_f} \rangle = {\cal F}^{\dagger} \arrowvert r_f 
\rangle$ is $4[1 - 4 O(\epsilon^2)/\pi^n]$. Hence, by using (9) the total
number of steps required to reach the target state is $N_s = O(\sqrt{\pi^n})$.

Now we show that the quantum search algorithm  over continuous variables
is robust to some extent. Instead of the Fourier transform ${\cal F}$ if we
replace it by a generalised Fourier transform (GFT) in the search operator
${\cal C}$, still the algorithm works, {\it i.e.}, we do get a square root 
reduction in the number of steps. We define a generalised Fourier transform 
as an active operation in the position basis $\arrowvert x \rangle$ as
\begin{eqnarray}
&& {\cal F}^{(\theta)} \arrowvert x \rangle = 
({i \over \pi \sin \theta})^{n/2} \nonumber \\
&& \times\int dy\, exp \bigg[-{i \over \sin \theta} 
[(x^2 + y^2) \cos \theta - 2xy] \bigg] \arrowvert y \rangle\;.
\end{eqnarray}
The GFT with a flexible angle $\theta$ gives a physical change of the basis
$\arrowvert x \rangle$ by any desired amount \cite{vn}. Here, it should be 
mentioned that $\theta > \arcsin ({1}/{\pi})$, since for smaller values 
of $\theta$ the assumption $\sin^n \theta \gg 1/\pi^n $ does not hold. 
The GFT for $\theta = 2 \pi m$, $m$ being an integer, corresponds
to no change of basis. The GFT for $\theta =\pi/ 2$ corresponds to
the Fourier transform defined in (1) (up to a constant phase shift equal 
to $n \pi/ 4$, $n$ being the number of qunats). If we apply GFT to an
initial basis $\arrowvert x_i \rangle$ then by probability rules of 
quantum theory we have to perform at least $O[{(\pi\sin\theta)}^n]$ number 
of trials to reach a target state $\arrowvert x_f \rangle$. We will prove 
that the generalised search operator acting on continuous variables will 
take $O[\sqrt{(\pi\sin\theta)^n}]$ steps to reach the final state.

The search operator with this GFT takes the form
\begin{equation}
{\cal C}^{(\theta)} = - I_{x_i}\, {{\cal F}^{(\theta)}}^{\dagger} 
I_{x_f} {\cal F}^{(\theta)}\;.
\end{equation}
We can see that the action of the generalised search operator on the 
initial state $\arrowvert x_i \rangle$ is given by
\begin{eqnarray}
&&{\cal C}^{(\theta)} \arrowvert x_i \rangle = 
\bigl(1 - {4 \over (\pi \sin \theta)^n} \bigr) \arrowvert x_i \rangle
+ 2 \sqrt{i^n \over {(\pi \sin \theta)}^n} \\
&&\times\!\int_{x_{f1}}^{x_{f2}} \!dx_f'\exp 
\bigl\{-{i \over \sin \theta} [(x_i^2 + x_f'^2) \cos \theta - 2x_ix_f] 
\bigr\} {\cal F}^{\dagger} \arrowvert x_f '\rangle\,. \nonumber
\end{eqnarray}
Similarly, the action of the generalised search operator 
${\cal C}^{(\theta)}$ on $\arrowvert \tilde{x} \rangle_f$ can be 
calculated. It is given by 
\begin{eqnarray}
&& {\cal C}^{(\theta)} \arrowvert \tilde{x} \rangle_f = 
\arrowvert \tilde{x} \rangle_f -2 
\sqrt{ {(-i)^n \over (\pi \sin \theta)^n } } \\
&&\times \int_{x_{i1}}^{x_{i2}} dx_i' \exp 
\bigl\{{i \over \sin \theta} [(x_i'^2 + x_f^2) \cos \theta - 2x_i'.x_f] 
\bigr\} \arrowvert x_i '\rangle\;. \nonumber
\end{eqnarray}
It can be seen that the generalised search operator creates linear
superposition of qunat states in the search process.  Now, we can
calculate the Fubini-Study distances to know how many steps are needed 
to reach the target state. The  shortest distance between the states 
$\arrowvert x_i \rangle$  and $\arrowvert \tilde{x}_f \rangle$ is
$d^2(\arrowvert x_i \rangle,\arrowvert \tilde{x}_f \rangle) =
4[1-1/(\pi\sin\theta)^n ]$. Notice that single application of the search 
operator ${\cal C}^{(\theta)}$ moves the initial state by a distance 
given by $d^2(\arrowvert x_i \rangle,\arrowvert x_i^{(1)} \rangle) =
({32 / \pi \sin \theta})^n$. Therefore, to travel a shortest distance
$d(\arrowvert x_i \rangle, \arrowvert \tilde{x}_f \rangle)$ we need
$N_s \approx O[\sqrt {(\pi \sin \theta)^n }]$ number of steps.

Thus, using a generalised Fourier transform we have proved that there is 
a square root reduction in the number of steps working with continuous 
variables.  As expected for an angle $\theta = \pi/ 2$ we get back 
the original result with the search operator ${\cal C}$. This result 
is similar to the recent result of Grover \cite{lgr}, where the search 
algorithm for qubits has been generalised for arbitrary unitary 
transformations.

In conclusion, we have for the first time provided an efficient algorithm 
such as quantum searching to be implemented on a quantum computer with 
continuous variables. The key elements in this generalisation are the Fourier 
transformation and inversion operators which constitute the search operator 
for qunats in an infinite dimensional Hilbert space. We find that a square 
root speed up is possible with quantum computers based on qunats. Also, the 
continuous variable search is possible with almost any Fourier 
transformation. This may be practically implemented for any large data 
base search using linear and non-linear optical devices with the role for
qunats being played by electromagnetic fields. It may well be that for 
large data base searches it is beneficial to use continuous quantum variables.

\vskip 0.1truein

\noindent
AKP and SLB acknowledge financial support from EPSRC.

\end{document}